\documentclass[aps,prl,showpacs,superscriptaddress,twocolumn]{revtex4-2}
\usepackage[colorlinks=true,letterpaper=true,pdfstartview=FitV,linkcolor=blue, citecolor=blue,urlcolor=blue,pdftitle=true]{hyperref}
\usepackage{graphicx}
\usepackage{amsmath}
\usepackage{bm}
\usepackage{pifont}
\usepackage{amssymb}
\usepackage{xcolor}
\usepackage{soul}

\begin{document}
\title{Spin Seebeck Effect in Normal-Metal--Chiral-Insulator Heterostructure}

\author{Jiayan Zhang}
\affiliation{Key Laboratory of Advanced Optoelectronic Quantum Architecture and Measurement, Ministry of Education, Beijing Institute of Technology, Beijing 100081, China}

\author{Gaoyang Li}
\affiliation{School of Physics $\&$ Information Science, Shaanxi University of Science $\&$ Technology, Xi'an 710021, China}

\author{Gaomin Tang}
\email[]{gmtang@gscaep.ac.cn}
\affiliation{Graduate School of China Academy of Engineering Physics, Beijing 100193, China}

\author{Yanxia Xing}
\email[]{xingyanxia@bit.edu.cn}
\affiliation{Key Laboratory of Advanced Optoelectronic Quantum Architecture and Measurement, Ministry of Education, Beijing Institute of Technology, Beijing 100081, China}

\begin{abstract}
Phonons can carry angular momentum and exhibit chirality through the circular polarization of atomic motion.
This enables a phonon-mediated spin Seebeck effect (SSE) via the conversion of phonon angular momentum into electron spin angular momentum.
In this Letter, we develop a theoretical framework for calculating the spin current in a normal-metal--chiral-insulator (NM--CI) heterostructure within the nonequilibrium Green's function formalism.
We discuss the influence of (i) the thermal bias across the NM--CI interface, (ii) the 
chemical potential of the NM, and (iii) the modification of the interfacial on-site 
potential on the spin transport properties.
We identify two characteristic nonlinear spin-transport phenomena: negative differential SSE and spin-current rectification.
The negative differential SSE arises from the competition between the thermal bias and the density of thermally excited electrons.
Spin-current rectification suggests the possibility of realizing a thermally controlled spin diode.
We also find that the spin-transport behavior is closely associated with an effective interfacial spectral density.
This work suggests a novel route toward thermally controlled spintronic devices using chiral phonons. 
\end{abstract}

\maketitle

\textit{Introduction}---A central goal of spintronics is to control and manipulate spin currents---the transport of electron spin angular momentum---in solid-state systems\,\cite{Igor2004, Bader2010, Supriyo2008}.
In addition to conduction electrons that carry spin angular momentum in metals and semiconductors, spin waves (magnons), the collective excitations of localized magnetic moments, can also transfer spin angular momentum even in ferromagnetic insulators (FIs)\,\cite{kajiwara2010, Cornelissen2015}.
Magnon spin currents can be generated by microwave irradiation or by applying a temperature gradient across the FI; these mechanisms are known as spin pumping and the spin Seebeck effect (SSE)\,\cite{Heinrich2011, Matsuo2018, Ominato2025, Adachi2013, Uchida2008, Uchida2009, Uchida2010}, respectively.
The resulting spin current can be detected in an adjacent normal metal (NM) via the inverse spin Hall effect\,\cite{Uchida2010, Saitoh2006, Valenzuela2006, Kimura2007}.
Magnon spin transport has also been widely studied in FI--NM\,\cite{Mahfouzi2014, LGY2022, LGY2024, TGM2018, ZJS2017, SYH2016}, FI--superconductor\,\cite{Kato2019, Johnsen2021, Silaev2020}, antiferromagnet--NM\,\cite{Seki2015, Rezende2016, Ohnuma2013}, and altermagnet systems\,\cite{Cui2023, Hodt2024, Sun2023}.

\begin{figure}[t]
	\centering
	\includegraphics[scale=0.51]{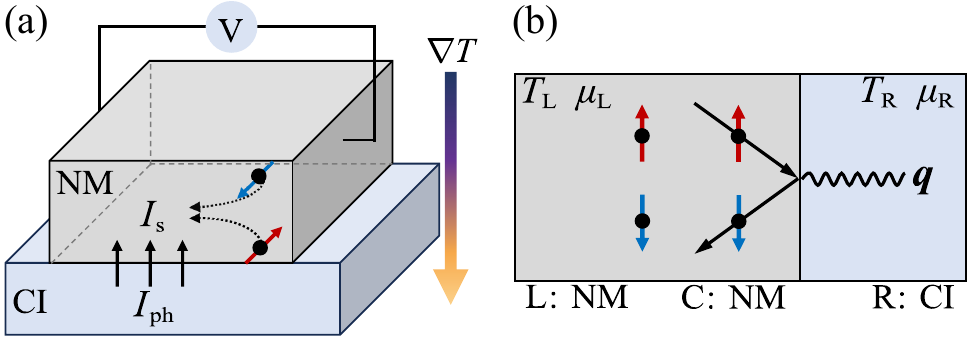}
	\caption{
		(a) Schematic of the experimental device. The temperature gradient $\nabla T$ is perpendicular to the NM--CI interface.
		The black arrows at the interface denote the conversion of the phonon angular momentum current $I_{\text{ph}}$ into the spin current $I_{\text{s}}$.
		Electrons in the NM are shown as black solid circles, with blue (red) arrows indicating the spin orientations.
		The black dotted lines with arrows are electron trajectories arising from spin-dependent scattering.
		(b) Schematic of the two-terminal representation of the NM--CI heterostructure, consisting of an NM left lead, an NM central region, and a CI phonon reservoir. $T_{\text{L}}$ $(T_{\text{R}})$ and $\mu_{\text{L}}$ $(\mu_{\text{R}})$ are the temperature and chemical potential of the left (right) lead, respectively.
		The black straight lines with arrows represent electrons before and after scattering, and the wavy line denotes phonon emission or absorption at the interface.   }
	\label{setup}
\end{figure}

Phonons, the quanta of lattice vibrations, are extensively discussed in the fields of superconductivity\,\cite{Bardeen1957}, thermal transport\,\cite{LNB2012, Yamamoto2006, WJS2006, Strohm2005, SL2006}, and nanophotonics\,\cite{Dai2014, Xu2014, Rivera2019}.
In many conventional treatments, phonons are described as linearly polarized lattice vibrations.
Recent theoretical and experimental studies have shown that atoms in solids can undergo circularly polarized motion, indicating that phonons can carry angular momentum and exhibit chirality\,\cite{ZLF2014, ZLF2015, Ueda2023, ZH2025, HYZ2018, IK2023, Juraschek2025, Juraschek2019, Kishine2025, CMQ2025, Ishito2023}.
Phonon chirality can give rise to a variety of intriguing transport phenomena, such as 
chirality-induced selectivity of phonon angular momenta\,\cite{Ohe2024}, phonon Edelstein 
effect\,\cite{Takehito2025}, chiral-phonon-induced spin transport\,\cite{QXX2025}, 
chiral-phonon diode effect\,\cite{CH2022}, orbital Seebeck effect\,\cite{Nabei2026}, and 
SSE due to spin-phonon conversion\,\cite{Kim2023, Funato2024, Naoki2025}.
However, nonlinear transport phenomena associated with the chiral-phonon-mediated SSE remain largely unexplored.
In this Letter, we employ the nonequilibrium Green's function formalism to investigate 
two characteristic nonlinear effects in this setting: the negative differential SSE and 
spin-current rectification.
The experimental setup consists of a chiral insulator (CI) adjacent to an NM, with a temperature gradient $\nabla T$ applied perpendicular to the NM--CI interface, as shown in Fig.\,\ref{setup}.
The thermally activated phonon angular momentum (phAM) current in the CI is converted into an electron spin current in the adjacent NM via interfacial spin-phonon conversion between the chiral phonons in the CI and the electrons in the NM.
Counterpropagating electrons with opposite spins in the NM are scattered toward the same side [black dotted line with arrows in Fig.\,\ref{setup}(a)].
As a result, an electron imbalance builds up between the two sides, leading to a measurable potential drop.

\textit{Model and Hamiltonian}---In this Letter, we develop a framework for calculating the spin current in an NM using the nonequilibrium Green’s function formalism\,\cite{Jauho2008, Stefanucci2025, Jishi2013}.
To model the SSE in the NM--CI system, we consider a two-terminal setup shown in Fig.\,\ref{setup}(b).
The total Hamiltonian reads
\begin{equation}
	\mathcal{H} = \mathcal{H}_\text{L} + \mathcal{H}_{\text{C}} + \mathcal{H}_\text{T} + \mathcal{H}_\text{ph} + \mathcal{H}_{e\text{-ph}}.
\end{equation}
Here $\mathcal{H}_{\text{L}}=\sum_{\bm{k}\sigma} \epsilon_{\bm{k}\sigma} 
b^\dagger_{\bm{k}\sigma} b_{\bm{k}\sigma}$ is the Hamiltonian of the left lead,
where $b^\dagger_{\bm{k}\sigma}(b_{\bm{k}\sigma})$ is the creation (annihilation) 
operator for an electron with wave vector $\bm{k}$ and spin $\sigma$.
We Fourier-transform $\mathcal{H}_{\text{L}}$ and obtain the corresponding lattice 
Hamiltonian as $H_{\text{L}}=2 \ell t \sum_{i\sigma} b^\dagger_{i\sigma} b_{i\sigma} - t 
\sum_{\langle ij \rangle \sigma} b^\dagger_{i\sigma} b_{j\sigma}$, where $\ell=1,2,3$ 
denotes the dimensionality and $t=21.768\,\text{meV}$ is the hopping parameter.
$\mathcal{H}_\text{C} = \sum_{n\sigma}\varepsilon_{n\sigma}c^\dagger_{n\sigma}c_{n\sigma}$ is the Hamiltonian of the central region,
where $c^\dagger_{n\sigma} (c_{n\sigma})$ creates (annihilates) an electron with energy $\varepsilon_{n\sigma}$.
$\mathcal{H}_{\text{T}}=\sum_{n\bm{k}\sigma}(t_{n\bm{k}\sigma}c^\dagger_{n\sigma} 
b_{\bm{k}\sigma} + \text{h.c.})$ describes the coupling between the left lead and the 
central region, where the tunneling matrix element is set to 
$t_{n\bm{k}\sigma}=t=21.768\,\text{meV}$ in our calculations.
$\mathcal{H}_\text{ph} = \sum_{\bm{q} \lambda}\hbar\omega_{\bm{q}\lambda} ( a^\dagger_{\bm{q}\lambda}a_ {\bm{q}\lambda} + 1/2 )$ is the Hamiltonian of the chiral phonons in the right lead,
where $a^\dagger_{\bm{q}\lambda} (a_ {\bm{q}\lambda})$ is the phonon creation (annihilation) operator with wave vector $\bm{q}$ and polarization $\lambda$, and $\lambda=\pm$ labels the left- and right-circularly polarized modes.
By considering the spin-microrotation coupling between electron spins in the NM and chiral phonons in the CI, the electron-phonon interaction $\mathcal{H}_{e\text{-ph}}$ is written as\,\cite{Funato2024}
\begin{equation}\label{H_eph}
	\mathcal{H}_{e\text{-ph}} = - \sum_{n\bm{q}\lambda} \Big(
	\mathcal{J}_{\bm{q}n}   d^\dagger_{\bm{q}\lambda} 
	c^\dagger_{n\downarrow}c_{n\uparrow} + 
	\mathcal{J}_{\bm{q}n}^\ast   d_{\bm{q}\lambda} c^\dagger_{n\uparrow}c_{n\downarrow} 
	\Big) .
\end{equation}
Here $\mathcal{J}_{\bm{q}n}$ denotes the electron-phonon coupling strength.
$d^\dagger_{\bm{q}\lambda} = \eta_{\bm{q}\lambda} \big(a_{\bm{q}\lambda} - a^\dagger_{-\bm{q}\bar{\lambda}} \big)$ and $d_{\bm{q}\lambda} =  \eta^\ast_{\bm{q}\lambda}  \big(a^\dagger_{\bm{q}\lambda} - a_{-\bm{q}\bar{\lambda}} \big)$ are the vorticity creation and annihilation operators with $\bar{\lambda}=-\lambda$,
where $\eta_{\bm{q}\lambda}=\sqrt{\hbar\omega_{\bm{q}\lambda} / (2\rho V)} \big[ ( \bm{q}\times \bm{e}_{\bm{q}\lambda} )_x + i ( \bm{q}\times \bm{e}_{\bm{q}\lambda} )_y \big] $ characterizes the propagation and polarization of chiral phonons\,\cite{Funato2024}.
The interaction Hamiltonian in Eq.\,(\ref{H_eph}) illustrates the following scattering process:
a spin-up (down) electron is scattered into a spin-down (up) state, accompanied by an increase (decrease) in phonon vorticity.
This process is crucial for spin-phonon conversion and thus for the generation of spin currents in the NM.

\textit{Derivation of the spin-current expression}---The system carries no net charge current because the right lead is insulating.
However, the system can carry a spin current due to the interfacial spin-phonon conversion.
The spin current in the left lead is defined as
\begin{equation}\label{Is_defi}
	I_\text{s} = \frac{\hbar}{2e} \big (I^e_{\uparrow} - I^e_{\downarrow} \big),
\end{equation} 
where the spin-dependent charge current $I^e_{\sigma}$ is\,\cite{Meir1992,Jauho2008}
\begin{equation}
	I^e_{\sigma} = \frac{e}{\hbar} \int \frac{dE}{2\pi}  \text{Tr}  \big[ 
	\textbf{G}^>_{\sigma}(E)  \bm{\Sigma}^<_{\text{L}\sigma}(E) - 
	\textbf{G}^<_{\sigma}(E)  \bm{\Sigma}^>_{\text{L}\sigma}(E)     \big]  .
\end{equation}
Boldface Latin and Greek letters denote matrices whose indices label lattice sites.
$\textbf{G}^{>/<}_\sigma$ are the greater and lesser electron Green's functions of the central region.
The retarded self-energy due to the left lead is $\bm{\Sigma}^r_{\text{L}\sigma}(E) = \textbf{H}^\dagger_{\text{T}} \textbf{g}^r_{\text{L}\sigma}(E) \textbf{H}_{\text{T}}$, where $\textbf{H}_{\text{T}}$ is the tunneling matrix between the left lead and the central region, and the left-lead surface Green's function $\textbf{g}^r_{\text{L}\sigma}$ is calculated from the left-lead lattice Hamiltonian $H_{\text{L}}$ using the iterative scheme\,\cite{Sancho1984, Sancho1985}.
The linewidth function of the left lead is $\bm{\Gamma}_{\text{L}\sigma}(E) = i\big[\bm{\Sigma}^r_{\text{L}\sigma}(E) - \bm{\Sigma}^a_{\text{L}\sigma}(E)\big] $.
For the left lead, $\bm{\Sigma}^{>}_{\text{L}\sigma}(E) = i \big[ f_{\text{L}}(E) - 1 \big] \bm{\Gamma}_{\text{L}\sigma}(E)$ 
and $\bm{\Sigma}^{<}_{\text{L}\sigma}(E) = i f_{\text{L}}(E) \bm{\Gamma}_{\text{L}\sigma}(E)$ 
where $f_{\text{L}}(E)=  1 \big/ \big[ e^{(E-\mu_{\text{L}}) / (k_\text{B}T_\text{L})}+1 \big]$ is the Fermi--Dirac distribution.

The spin current can be calculated once we obtain the central-region Green’s functions.
The greater/lesser and retarded Green's functions in the central region are given by [see the Supplemental Material \cite{SuppMat} for details of the derivation]
\begin{align}
	\hspace{-0.2cm} \textbf{G}^{>/<}_{\sigma}(E) & = \textbf{G}^{r}_{\sigma}(E)  \big[  \bm{\Sigma}^{>/<}_{\text{L}\sigma}(E) + \overline{\boldsymbol{\Sigma}}^{>/<}_{\text{R}\sigma}(E) \big] \textbf{G}^{a}_{\sigma}(E)  , \label{keldysh} \\
	\textbf{G}^{r}_{\sigma}(E) & = \big[ E\textbf{I} - \textbf{H}_{\text{C}} - \boldsymbol{\Sigma}^{r}_{\text{L}\sigma}(E) - \overline{\boldsymbol{\Sigma}}^{r}_{\text{R}\sigma}(E) \big]^{-1} . \label{dyson} 
\end{align}	
Here, $\textbf{I}$ is the unit matrix.
$\bm{\Sigma}_{\text{L}\sigma}$ is the left-lead self-energy.
$\overline{\bm{\Sigma}}_{\text{R}\sigma}$ is the phononic self-energy and is expressed as\,\cite{SuppMat}
\begin{align}
	\overline{\Sigma}^{>/<}_{\text{R}\uparrow,ij}(E) & = i \int \frac{d\omega}{2\pi} G^{>/<}_{\downarrow,ij}(E-\hbar\omega) \Sigma^{>/<}_{\text{R},ij}(\omega) , \label{selfenergy_1}  \\
	\overline{\Sigma}^{>/<}_{\text{R}\downarrow,ij}(E) & = i \int \frac{d\omega}{2\pi} G^{>/<}_{\uparrow,ij}(E+\hbar\omega)  \Sigma^{</>}_{\text{R},ji}(\omega)   ,  \label{selfenergy_2}  \\
	\overline{\Sigma}^{r}_{\text{R}\uparrow,ij}(E) & = i \int \frac{d\omega}{2\pi} \Big[ G^{r}_{\downarrow,ij}(E-\hbar\omega) \Sigma^{<}_{\text{R},ij}(\omega) \nonumber\\
	& \qquad\qquad \ + G^{>}_{\downarrow,ij}(E-\hbar\omega) \Sigma^{r}_{\text{R},ij}(\omega)  \Big] , \label{selfenergy_3}  \\
	\overline{\Sigma}^{r}_{\text{R}\downarrow,ij}(E)   & = i \int \frac{d\omega}{2\pi} \Big[G^{r}_{\uparrow,ij}(E+\hbar\omega)  \Sigma^{<}_{\text{R},ji}(\omega) \nonumber\\ 
	& \qquad\qquad \ + G^{<}_{\uparrow,ij}(E+\hbar\omega) \Sigma^{a}_{\text{R},ji}(\omega) \Big] . \label{selfenergy_4}
\end{align}
Here, the subscripts $i$ and $j$ label the matrix elements.
$\bm{\Sigma}_{\text{R}}$ is the right-lead self-energy, and its relations to the linewidth function $\bm{\Gamma}_{\text{R}}$ are $\bm{\Sigma}^{r}_{\text{R}}(\omega) = [\bm{\Sigma}^{a}_{\text{R}}(\omega)]^{\dagger} = -i\bm{\Gamma}_{\text{R}}(\omega)/2$,
$\boldsymbol{\Sigma}^{>}_{\text{R}}(\omega)=-i[n_{\text{R}}(\omega)+1] \boldsymbol{\Gamma}_{\text{R}}(\omega)$,
and $\boldsymbol{\Sigma}^{<}_{\text{R}}(\omega) = -in_{\text{R}}(\omega) \boldsymbol{\Gamma}_{\text{R}}(\omega)$ where $n_{\text{R}}(\omega) =  1 \big/ \big[ e^{(\hbar\omega)/(k_\text{B}T_\text{R})}-1 \big]$ is the Bose--Einstein distribution.
In the presence of phonon chirality, phonon branches with opposite circular polarizations 
exhibit nondegenerate dispersions. Under a thermal bias, the resulting imbalance in their 
occupations can give rise to a net phonon angular momentum current. By contrast, in the 
nonchiral limit, the angular-momentum contributions from modes with opposite circular 
polarizations cancel. Since the present model does not microscopically resolve the 
individual chiral branches, we represent the net chiral-phonon contribution of the right 
reservoir by a super-Ohmic spectral density\,\cite{Weissbook, Thorwart2005, TQS2022}
\begin{equation}
	J_{\text{R}}(\omega) = \alpha \hbar\frac{\omega^3}{\omega_0^{2}} e^{-\omega/\omega_0} ,   
\end{equation}
where the dimensionless parameter $\alpha$ describes the dissipation into the reservoir and $\hbar\omega_0$ is the cutoff energy.
The spectral density $J_{\text{R}}$ describes the coupling strength between the phonons with energy $\hbar\omega$ in the right reservoir and the electrons in the central region.
Using this spectral density, the right-lead linewidth function is taken to be $\bm{\Gamma}_{\text{R}}(\omega) = J_{\text{R}}(\omega) \textbf{I}$.

Because the net charge current vanishes, $I^e_{\uparrow} + I^e_{\downarrow} = 0$, and thus $I^e_{\uparrow} = -I^e_{\downarrow}$.
For simplicity, we focus on the spin-up contribution, i.e., $I_{\text{s}} = (\hbar/e)I^e_{\uparrow}$.
Using Eq.\,(\ref{keldysh}), we obtain the expression for the spin current
\begin{equation}\label{Is}
	I_{\text{s}} = \int \frac{dE}{2\pi} \text{Tr} \Big\{ 
	\textbf{G}^r_{\uparrow} \Big[
	if_{\text{L}} \overline{\boldsymbol{\Sigma}}^{>}_{\text{R}\uparrow} + i(1-f_\text{L}) \overline{\boldsymbol{\Sigma}}^{<}_{\text{R}\uparrow}
	\Big] \textbf{G}^a_{\uparrow} \boldsymbol{\Gamma}_{\text{L}\uparrow} \Big\} .
\end{equation}
The Green's functions in Eqs.\,(\ref{keldysh}) and (\ref{dyson}), together with the phonon self-energies in Eqs.\,(\ref{selfenergy_1})-(\ref{selfenergy_4}), form a closed set of equations and are solved self-consistently until convergence is reached.
The left-lead spin current $I_{\text{s}}$ is then calculated using Eq.\,(\ref{Is}).
In this Letter, we focus only on the spin current in the left lead.
In the Supplemental Material, we derive the expression for the phAM current in the right lead and verify the current conservation\,\cite{SuppMat}.

\textit{Negative differential SSE}---We consider a two-terminal setup to simulate the SSE in the NM--CI system, as shown in Fig.\,\ref{setup}(b).
$T_{\text{L}}$ and $T_{\text{R}}$ denote the temperatures of the left and right leads, respectively. 
Thus the thermal bias is $\Delta T = T_{\text{L}} - T_{\text{R}}$.
We further define $T_0 = (T_\text{L} + T_{\text{R}})/2$ and then $T_{\text{L}}=T_0 + \Delta T/2$ and $T_{\text{R}}=T_0 - \Delta T/2$.
The chemical potential of the left lead $\mu_{\text{L}}$ is tunable, whereas we set $\mu_{\text{R}}=0$.
We introduce a shift $\varepsilon_d$ of the interfacial on-site potential to examine its influence on spin transport.
The phenomena discussed in this work do not qualitatively depend on the 
dimensionality of the system. Numerical tests show that the spin current is independent 
of the central-region length along the transport direction.
Therefore, for simplicity, the central region is modeled as a quantum dot, whose 
Hamiltonian is given by $H_{\text{dot}} = 2t \sum_{\sigma} c^\dagger_{\sigma} c_{\sigma}$ 
with $t = 21.768 \, \rm meV$.
We also present the spin-current results for different central-region sizes in the 
Supplemental Material\,\cite{SuppMat}.

\begin{figure}[t]
	\centering
	\includegraphics[scale=0.36]{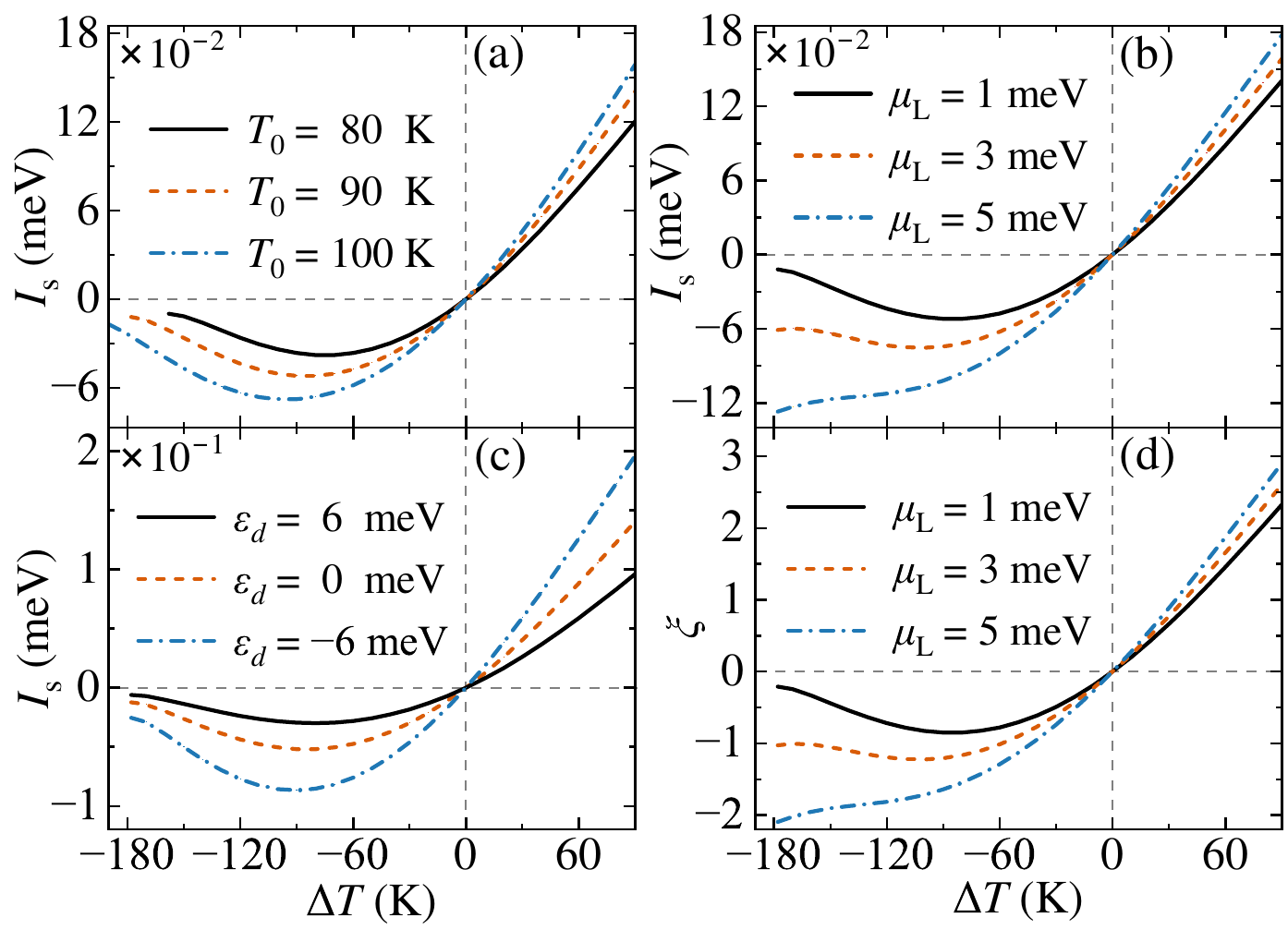}
	\caption{Spin current $I_{\text{s}}$ as a function of the thermal bias $\Delta T$ 
		(a) for different $T_0$ with $\mu_{\text{L}}=1\,\rm meV$ and $\varepsilon_d=0$, 
		(b) for different $\mu_{\text{L}}$ with $T_0=90\,\rm K$ and $\varepsilon_d=0$, 
		and (c) for different $\varepsilon_d$ with $\mu_{\text{L}}=1\,\rm meV$ and $T_0=90 \,\rm K$.
		(d) $\xi$ plotted as a function of $\Delta T$ for different $\mu_{\text{L}}$ with $T_0=90\,\rm K$ and $\varepsilon_d=0$. The other parameters are $\alpha=5$ and $\hbar\omega_0=7\,\rm meV$.	}
	\label{Fig2}
\end{figure}

We first study the dependence of the spin current $I_{\text{s}}$ on the thermal bias $\Delta T$ at different average temperatures $T_0$,
with $\mu_{\text{L}}=1$\,meV and $\varepsilon_d=0$, as shown in Fig.\,\ref{Fig2}(a).
For $\Delta T > 0$, the spin current $I_{\text{s}}$ increases monotonically with $\Delta T$.
In addition, the overall magnitude of $I_{\text{s}}$ increases as $T_0$ increases.
For $\Delta T < 0$, the magnitude $|I_{\text{s}}|$ initially increases with $|\Delta T|$, but begins to decrease once $|\Delta T|$ exceeds a critical value.
This counterintuitive regime, in which the magnitude of the spin current decreases as the magnitude of the thermal bias increases, is referred to as the negative differential SSE.
Negative differential SSE has been reported in phonon thermal transport\,\cite{LBW2004} and magnon spin transport\,\cite{RJ2013}.
Here we show that this phenomenon can also occur in the NM--CI heterostructure.

The negative differential SSE can be explained in terms of the competition between the thermal bias and the left-lead electron density.
For $\Delta T > 0$, increasing $\Delta T$---that is, raising $T_{\text{L}}$ and lowering $T_{\text{R}}$---increases the population of thermally excited electrons in the left lead and promotes spin-phonon conversion at the interface, thereby enhancing the spin current.
Consequently, $I_{\text{s}}$ increases monotonically as $\Delta T$ increases.
For $\Delta T <0$, increasing $|\Delta T|$ enhances the thermal driving experienced by electrons in the central region.
Meanwhile, in this process, $T_{\text{L}}$ decreases and the thermal excitation of electrons is suppressed, which weakens the spin-phonon conversion and reduces the spin current.
Thus the competition between these two effects gives rise to negative differential SSE at lower $T_{\text{L}}$.

We further examine the influence of the left-lead chemical potential $\mu_{\text{L}}$ on the spin transport, as shown in Fig.\,\ref{Fig2}(b).
For $\mu_{\text{L}}=1$\,meV, the spin current still exhibits pronounced negative differential SSE.
As $\mu_{\text{L}}$ increases, the negative differential effect becomes less pronounced at $\mu_{\text{L}}=3$\,meV and eventually disappears at $\mu_{\text{L}}=5$\,meV.
Since the decrease in thermally excited electron density due to lowering $T_{\text{L}}$ is compensated by the increase in electron density caused by raising the chemical potential, the negative differential SSE disappears for a higher chemical potential.

As demonstrated in studies of the magnon-mediated SSE\,\cite{WHL2014, LWW2016, LWY2021}, inserting a thin NiO or MoS$_2$ layer at the YIG--Pt interface can significantly enhance the spin current.
Motivated by these experiments, we examine one possible electronic consequence of an 
atomically thin interfacial modification by shifting the on-site potential of the lattice 
site adjacent to the right reservoir, $2t \rightarrow 2t + \varepsilon_d$.
This level shift is motivated by the interface-dipole mechanism demonstrated 
experimentally in Ref.\,\cite{Yajima2015}, in which a charged atomic-scale interlayer and 
its compensating screening charge generate an electrostatic potential step.
Figure\,\ref{Fig2}(c) presents $I_\text{s}$ for different $\varepsilon_d$ at $T_0=90$\,K and $\mu_{\text{L}}=1$\,meV.
The spin current increases (decreases) for $\varepsilon_d <0 \, (\varepsilon_d >0)$, demonstrating that a shift of the interfacial on-site potential can effectively tune spin transport in the NM–CI heterostructure.
In the present calculation, we isolate the influence of an interfacial on-site potential 
shift on spin transport, while the hopping amplitudes, electron-phonon coupling, and 
phonon spectral density are held fixed.
Figure\,\ref{Fig2}(c) should therefore be interpreted as a one-parameter sensitivity analysis rather than as a complete microscopic description of an inserted layer.
This approximation is most appropriate for an atomically thin interlayer whose induced electrostatic perturbation is localized within the interfacial region\,\cite{Tung2001}.
For a thicker interlayer, an explicit interlayer Hamiltonian together with modified 
hopping parameters would be required.

\begin{figure}[t]
	\centering
	\includegraphics[scale=0.355]{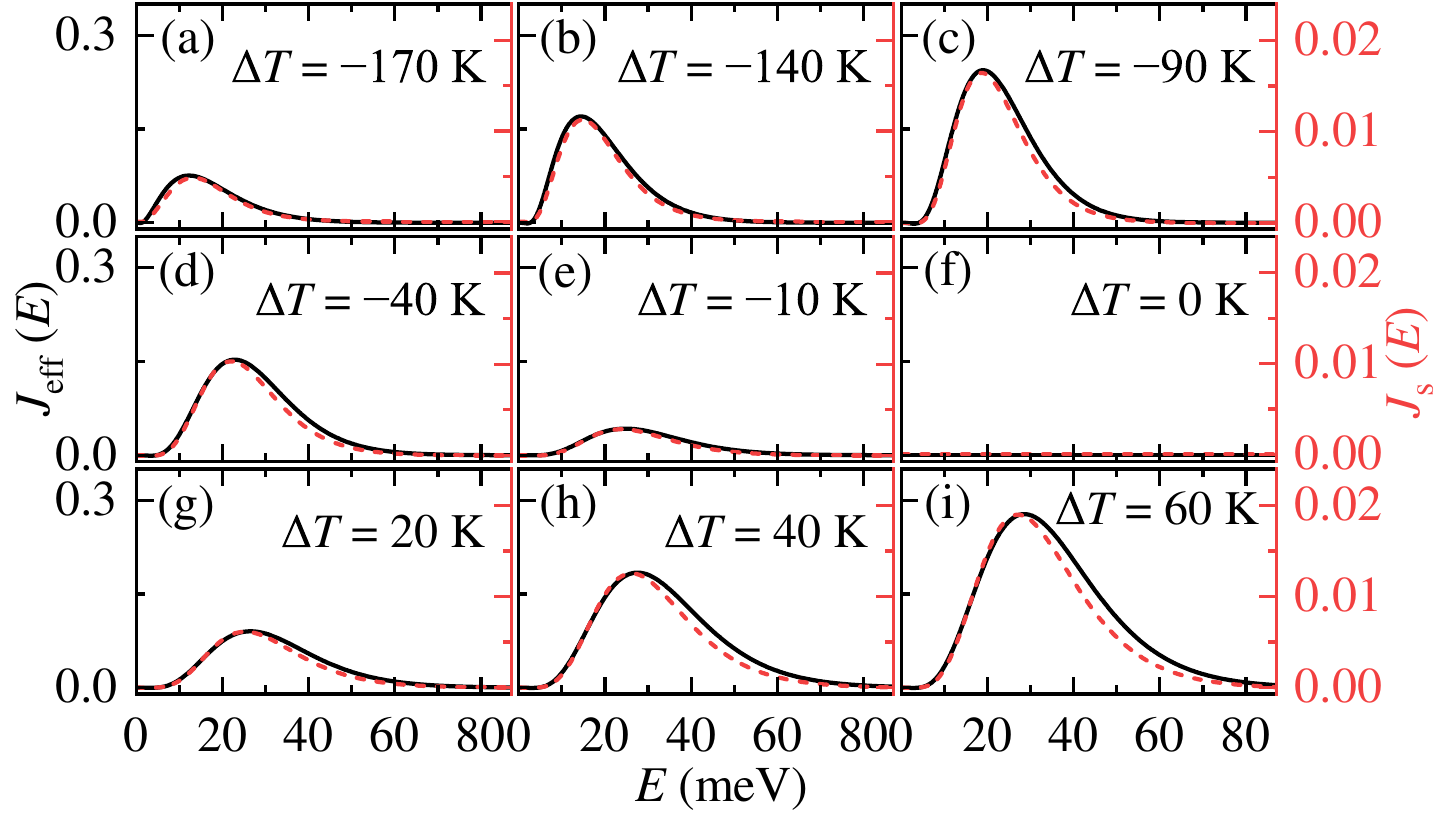}
	\caption{Effective spectral density $J_{\text{eff}}(E)$ and spin current density $J_{\text{s}}(E)$ as a function of energy $E$ at different $\Delta T$ for $\mu_{\text{L}}=1\,\rm meV$.
		Other parameters: $T_0=90\,\rm K$, $\varepsilon_d=0$, $\alpha=5$, and $\hbar\omega_0=7\,\rm meV$. }
	\label{Fig3}
\end{figure}

By comparing the spin-current expression in Eq.\,(\ref{Is}) with the $\rm 
Landauer\text{--}B\ddot{u}ttiker$ formula\,\cite{Dattabook}, we can define the effective 
spectral density $J_{\text{eff}}$ as\,\cite{Sergueev2002} 
\begin{equation}\label{Jeff}
	J_{\text{eff}}(E) = \text{Tr} \Big\{ if_{\text{L}}(E) \overline{\boldsymbol{\Sigma}}^{>}_{\text{R}\uparrow}(E)  + i \big[1-f_\text{L}(E) \big] \overline{\boldsymbol{\Sigma}}^{<}_{\text{R}\uparrow}(E) \Big\} .
\end{equation}
The effective spectral density $J_{\text{eff}}$ can be interpreted as the effective coupling strength between the NM and the CI.
This interpretation follows from the fact that the phonon self-energy in $J_{\text{eff}}$ is written in the simplified form $\overline{\Sigma} \sim G\Sigma_{\text{R}}$, which represents the coupling between the central region and the right lead.
Accordingly, we define $\xi$ as
\begin{equation}\label{xi}
	\xi = \int J_{\text{eff}}(E) \, dE  ,
\end{equation}
which represents the overall contribution of the effective interfacial coupling.
Below, we show that the spin-transport behavior is closely related to $\xi$ and $J_{\text{eff}}$.

Figure\,\ref{Fig2}(d) shows $\xi$ as a function of $\Delta T$ for different $\mu_{\text{L}}$ at $T_0=90$\,K and $\varepsilon_d=0$.
A comparison with $I_\text{s}$ in Fig.\,\ref{Fig2}(b) reveals a strong correlation between the spin current and $\xi$.
To elucidate the spin transport properties in detail, we plot the effective spectral density $J_{\text{eff}}$ and the spin-current spectrum $J_{\text{s}}$ at different values of $\Delta T$ for $\mu_{\text{L}}=1$\,meV, as shown in Fig.\,\ref{Fig3}.
The spin-current spectrum $J_{\text{s}}$ is defined as the integrand in Eq.\,(\ref{Is}).
As $\Delta T$ decreases from 0 to $-170\, \rm K$, equivalently as $|\Delta T|$ increases, both $J_{\rm eff}$ and $J_{\rm s}$ initially increase [Figs.\,\ref{Fig3}(c)--\ref{Fig3}(f)], reach their maxima at $\Delta T = -90 \, \rm K$ [Fig.\,\ref{Fig3}(c)], and then decrease [Figs.\,\ref{Fig3}(a)--\ref{Fig3}(c)]. 
This trend is fully consistent with the variation of $I_{\rm s}$ in Fig.\,\ref{Fig2}(b) for $\mu_{\text{L}}=1 \, \rm meV$.
Thus, as shown in Figs.\,\ref{Fig2}(d) and \ref{Fig3}, the strong correlations between $\xi$ and $I_\text{s}$, and between $J_\text{eff}$ and $J_\text{s}$, illustrate the crucial role of the effective interfacial spectral density in determining the spin-transport behavior in the NM--CI heterostructure.

\begin{figure}[t]
	\centering
	\includegraphics[scale=0.38]{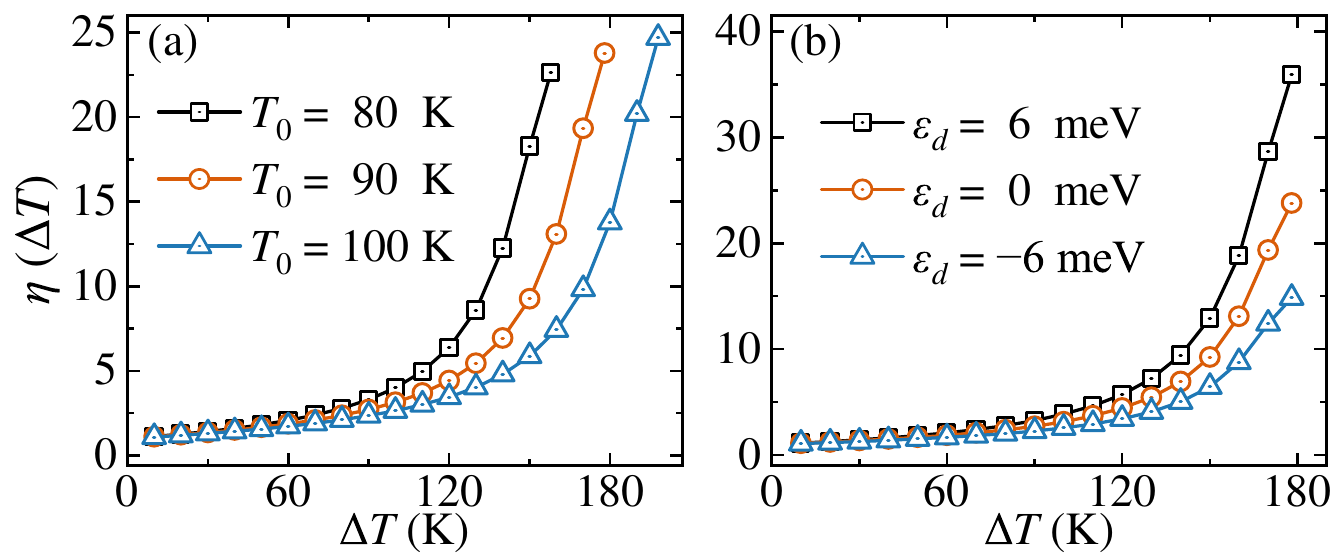}
	\caption{(a) Rectification ratio $\eta$ as a function of $\Delta T$ for different $T_0$ with $\mu_{\text{L}}=1\,\rm meV$ and $\varepsilon_d=0$. (b) $\eta$ as a function of $\Delta T$ for different $\varepsilon_d$ with $T_0=90 \,\rm K$ and $\mu_{\text{L}}=1 \,\rm meV$.}
	\label{Fig4}
\end{figure}

\textit{Rectification effect and thermal spin diode}---As shown in Fig.\,\ref{Fig2}, in addition to the negative differential SSE discussed above, the spin current also exhibits rectification; that is, its magnitude is asymmetric under reversal of the thermal bias: $| I_{\text{s}}(\Delta T) | \neq | I_{\text{s}}(-\Delta T) |$.
The spin-current asymmetry stems from the structural asymmetry between the left and right leads,
and the resulting nonreciprocal spin transport can be used to realize a thermally controlled spin diode effect\,\cite{RJ2013_2}.
We can define a rectification ratio $\eta$ to quantify the rectification performance of the spin diode as
\begin{equation}
	\eta(\Delta T) = \left| \frac{I_{\text{s}}(\Delta T)}{I_{\text{s}}(-\Delta T)} 
	\right| , \quad \Delta T >0 .
\end{equation}
The ratio is widely used to quantify the performance of thermal\,\cite{LBW2004, RJ2013_3}, electronic\,\cite{CR2014, LZL2020}, and spin diodes\,\cite{Iovan2008}.
Figure\,\ref{Fig4}(a) presents the rectification ratio for different $T_0$ at $\mu_{\text{L}}=1 \, \rm meV$ and $\varepsilon_d=0$.
The rectification ratio increases with the thermal bias $\Delta T$ and reaches a maximum value of approximately 25.
Figure\,\ref{Fig4}(b) shows the rectification ratio for different $\varepsilon_d$ at $T_0=90 \,\rm K$ and $\mu_{\text{L}}=1 \, \rm meV$.
The results show that $\eta$ can be enhanced (diminished) for $\varepsilon_d>0 \, (\varepsilon_d<0)$, demonstrating that the interfacial on-site potential provides an effective means to tune the performance of the thermal spin diode.

\textit{Conclusion}---In summary, we investigate the nonlinear spin-transport properties of the chiral-phonon-mediated spin Seebeck effect.
We consider a two-terminal NM--CI setup and develop a framework for calculating the spin current within the nonequilibrium Green's function formalism.
We discuss the influence of (i) the temperature difference between the NM and the CI,
(ii) the chemical potential of the NM, and (iii) the modification of the interfacial on-site potential on the spin transport properties.
We identify two characteristic nonlinear spin-transport phenomena: negative differential SSE and spin-current rectification.
The negative differential effect arises from the competition between the thermal bias and 
the thermally excited electron density, and the rectification effect originates from the 
structural asymmetry of the two-terminal setup.
The spin-transport behavior is closely related to an effective interfacial spectral density $J_{\text{eff}}$ and its energy integral $\xi$.
Furthermore, we suggest that spin-current rectification can be exploited to realize a thermal spin diode.
We calculate the corresponding rectification ratio $\eta$ to characterize the diode efficiency.
This work provides a novel approach to realizing thermally controlled spintronic devices using chiral phonons.

\textit{Acknowledgments}---J.\,Z. and Y.\,X. are supported by the National Natural Science Foundation of China (Grant No. 12174023).
G.\,L. is supported by the Youth Innovation Team of Shaanxi Universities (Grant No. 25JP021).
G.\,T. is supported by the National Natural Science Foundation of China (Grants No. 12088101 and No. 12374048).

\bibliography{reference_main}

\end{document}


\title{Supplementary Materials for ``Spin Seebeck Effect in Normal-Metal--Chiral-Insulator Heterostructure'' }
	
\author{Jiayan Zhang}
\affiliation{Key Laboratory of Advanced Optoelectronic Quantum Architecture and Measurement, Ministry of Education, Beijing Institute of Technology, Beijing 100081, China}

\author{Gaoyang Li}
\affiliation{School of Physics $\&$ Information Science, Shaanxi University of Science $\&$ Technology, Xi'an 710021, China}

\author{Gaomin Tang}
\email[]{gmtang@gscaep.ac.cn}
\affiliation{Graduate School of China Academy of Engineering Physics, Beijing 100193, China}

\author{Yanxia Xing}
\email[]{xingyanxia@bit.edu.cn}
\affiliation{Key Laboratory of Advanced Optoelectronic Quantum Architecture and Measurement, Ministry of Education, Beijing Institute of Technology, Beijing 100081, China}

\maketitle
	
	
	
\tableofcontents

\section{S1. The expression of the spin current in the left lead}\label{Is_L}
In this section, we present the expression for the spin current in the left lead, defined as
\begin{equation}\label{Is1}
	I_{\text{s}} = \frac{\hbar}{2e} \left(I^e_{\uparrow} - I^e_{\downarrow} \right) ,
\end{equation}
where $I^e_{\sigma}$ is the spin-dependent charge current in the left lead.
From the spin-dependent electron-number operator in the left lead, 
$\mathcal{N}_{\text{L}\sigma} = \sum_{\bm{k}} b^\dagger_{\bm{k}\sigma} b_{\bm{k}\sigma}$, 
the charge current is
\begin{align}\label{lsc1}
	I^e_{\sigma}(t) & = -e \Big\langle \frac{d}{dt}\mathcal{N}_{\text{L}\sigma} \Big\rangle 
	= - \frac{ie}{\hbar} \big\langle \big[ \mathcal{H}, \mathcal{N}_{\text{L}\sigma} \big] \big\rangle \nonumber\\
	& =  -\frac{ie}{\hbar} \sum_{n\bm{k}} 
	\left[ t_{n\bm{k}\sigma} \big\langle c^\dagger_{n\sigma}(t) 
	b_{\bm{k}\sigma}(t)\big\rangle - 
	t^\ast_{nk\sigma} \big\langle  b^\dagger_{\bm{k}\sigma}(t) c_{n\sigma}(t) 
	\big\rangle  \right] .
\end{align}
We define the lesser mixed Green's functions between the left lead (L) and the central region (C) by
\begin{equation}\label{G<}
	G^<_{\sigma,\bm{k}n}(t,t') = \frac{i}{\hbar} \big\langle c^\dagger_{n\sigma}(t') 
	b_{\bm{k}\sigma}  (t) \big\rangle, \quad\quad\quad
	G^<_{\sigma,n\bm{k}}(t,t') = \frac{i}{\hbar} \big\langle  b^\dagger_{\bm{k}\sigma} 
	(t') c_{n\sigma}(t) \big\rangle ,   
\end{equation}
We have $G^<_{\sigma,\bm{k}n}(t,t') = -\big[G^<_{\sigma,n\bm{k}}(t',t) \big] ^\dagger$.
Then the spin-dependent electric current can be expressed as
\begin{equation}\label{I-G<}
	I^e_{\sigma}(t) = -e\sum_{n\bm{k}} t_{n\bm{k}\sigma}  \big[ G^<_{\sigma,\bm{k}n}(t,t)  + \text{h.c.} \big] .
\end{equation}
\begin{figure}[t]
	\centering
	\includegraphics[scale=0.50]{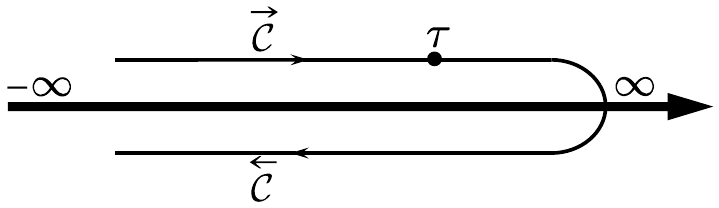}
	\caption{ Keldysh contour. The contour runs from $t=-\infty$ to $t=\infty$ along the forward branch $\overrightarrow{\mathcal{C}}$, and returns from $t=\infty$ to $t=-\infty$ along the backward branch $\overleftarrow{\mathcal{C}}$.	}
	\label{KeldyshC}
\end{figure}
To obtain $G^<_{\sigma,\bm{k}n}(t,t)$, we first define the contour-ordered Green's function as
\begin{equation}\label{Gtau}
	G_{\sigma,\bm{k}n}(\tau,\tau') = -\frac{i}{\hbar} \left\langle \mathcal{T}_c 
	b_{\bm{k}\sigma}(\tau) c^\dagger_{n\sigma}(\tau') \right\rangle 
	= -\frac{i}{\hbar} \left\langle \mathcal{T}_c \mathcal{S} b_{\bm{k}\sigma}(\tau) 
	c^\dagger_{n\sigma}(\tau')\right\rangle _0  . 
\end{equation}
Here $\tau$ and $\tau'$ are the variables on the Keldysh contour [see 
Fig.~\ref{KeldyshC}] and $\mathcal{T}_c$ is the contour-ordering operator.
$\langle \,\cdots \rangle$ denotes the ensemble average taken with respect to the coupled L-C system,
while $\langle \,\cdots \rangle_0$ is taken with respect to the decoupled L-C system
in which the left lead and the central region are each in their own thermal equilibrium.
The $\mathcal{S}$-matrix is defined as
\begin{equation}\label{S_mat}
	\mathcal{S} = \mathcal{T}_c \exp \left[ -\frac{i}{\hbar}\int_c d\tau \mathcal{H}_{\text{T}}(\tau) \right] = \sum_{n=0}^{\infty} \left(-\frac{i}{\hbar} \right)^n\frac{1}{n!} \int_cd\tau_1 \cdots \int_cd\tau_n \mathcal{T}_c \big[ \mathcal{H}_{\text{T}}(\tau_1)\cdots \mathcal{H}_{\text{T}}(\tau_n) \big] .
\end{equation}
Since expectation values containing an odd number of fermion operators vanish, we retain only even orders in $t_{n\bm{k}\sigma}$ in the expansion of $G_{\sigma,\bm{k}n}(\tau,\tau')$. At zeroth order, $G_{\sigma,\bm{k}n}^{(0)}(\tau,\tau')=0$. To second order, we obtain
\begin{equation}
	G_{\sigma,\bm{k}n}^{(2)}(\tau,\tau') = \sum_{m} \int_c d\tau_1 \big[t_{m\bm{k}\sigma}^\ast  g_{\bm{k}\sigma}(\tau,\tau_1)  g_{\sigma,mn}(\tau_1,\tau') \big] ,
\end{equation}
where we defined the following noninteracting Green's functions of the left lead and the central region
\begin{equation}
	g_{\bm{k}\sigma}(\tau,\tau_1) = -\frac{i}{\hbar} \big\langle \mathcal{T}_c 
	b_{\bm{k}\sigma}(\tau) b^\dagger_{\bm{k}\sigma}  (\tau_1) \big\rangle _0 , 
	\qquad\qquad 
	g_{\sigma,mn}(\tau_1,\tau') = -\frac{i}{\hbar} \left\langle \mathcal{T}_c c_{m\sigma}(\tau_1) c^\dagger_{n\sigma}(\tau')\right\rangle _0 .
\end{equation}
Summing contributions from all higher even orders, the  Green’s function becomes
\begin{equation}
	G_{\sigma,\bm{k}n}(\tau,\tau') = \sum_m \int_c d\tau_1 \big[ t_{m\bm{k}\sigma}^\ast g_{\bm{k}\sigma}(\tau,\tau_1) G_{\sigma,mn}(\tau_1,\tau') \big] .
\end{equation}
Applying the Langreth rules~\cite{Jauho2008} and taking the Fourier transform, we obtain
\begin{equation}
	G^<_{\sigma,\bm{k}n}(t-t') = \frac{1}{\hbar}\sum_m \int \frac{dE}{2\pi}  \big[ 
	g^r_{\bm{k}\sigma}(E) t_{\bm{k}m\sigma} G^<_{\sigma,mn}(E)  +  g^<_{\bm{k}\sigma}(E) t_{\bm{k}m\sigma} G^a_{\sigma,mn}(E) \big] e^{-iE(t-t')/\hbar} .
\end{equation}
Using
\begin{equation}
	G^{<,\ast}_{\sigma,nm}(E) = - G^<_{\sigma,mn}(E) ,\qquad\qquad      G^{a,\ast}_{\sigma,nm}(E) = G^r_{\sigma,mn}(E)  ,
\end{equation}
and defining the retarded ($r$), advanced ($a$) and lesser ($<$) self-energies of the left lead
\begin{equation}
	\Sigma^{r/a/<}_{\text{L}\sigma,mn}(E) = \sum_{\bm{k}} t_{m\bm{k}\sigma} \; g^{r/a/<}_{\bm{k}\sigma}(E) \; t_{\bm{k}n\sigma} ,
\end{equation}
we obtain the spin-dependent electric current in the left lead
\begin{equation}\label{I_e}
	I^e_{\sigma} = \frac{e}{\hbar} \int \frac{dE}{2\pi}  \text{Tr} \Big[  \textbf{G}^>_{\sigma}(E) \bm{\Sigma}^<_{\text{L}\sigma}(E) - \textbf{G}^<_{\sigma}(E) \bm{\Sigma}^>_{\text{L}\sigma}(E) \Big] .
\end{equation}
After obtaining the $I^e_{\sigma}$, the spin current is calculated using Eq.~(\ref{Is1}).

\section{S2. The Green's function of the central region}\label{G_c}
\begin{figure}[t]
	\centering
	\includegraphics[scale=0.50]{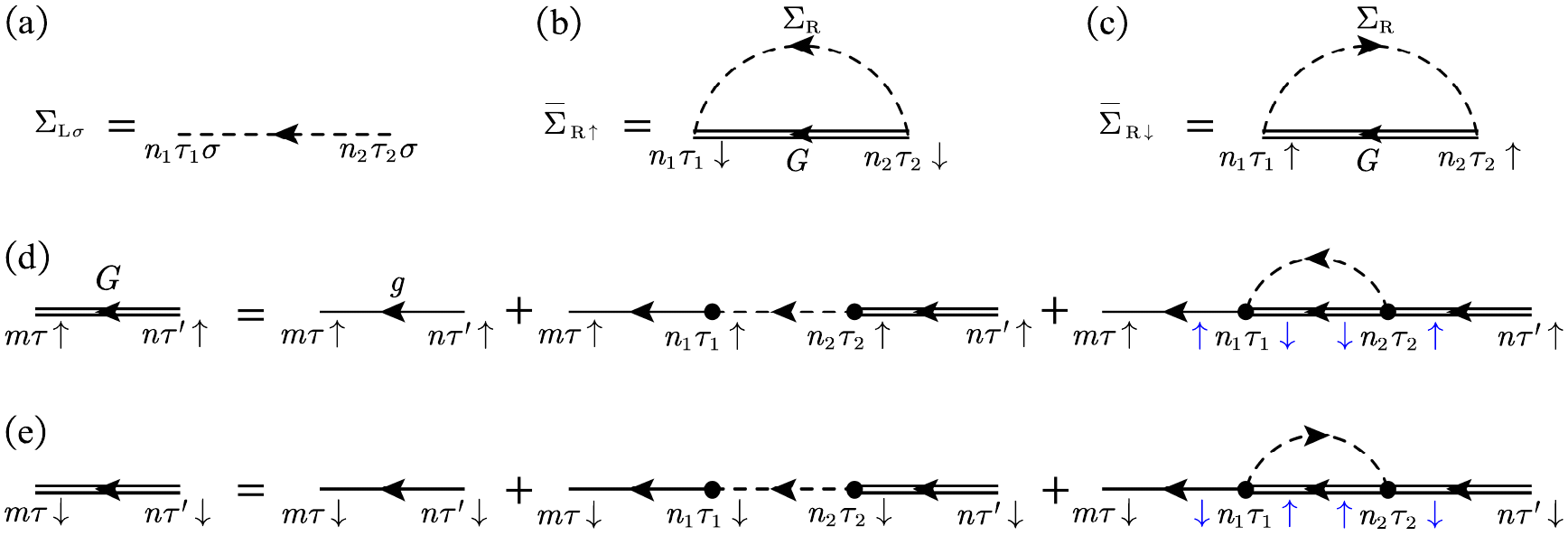}
	\caption{
		Feynman diagrams for the electron self-energy $\Sigma_{\text{L}\sigma}$, the phonon self-energy $\overline{\Sigma}_{\text{R} \sigma}$ and the central-region Green's functions $G_{\sigma,mn}$.
		A single solid line with an arrow represents the noninteracting Green’s function $g$,
		whereas a double solid line with an arrow represents the interacting Green’s function $G$.
		A straight dotted line with an arrow denotes the left-lead self-energy $\Sigma_{\text{L}\sigma}$
		and a curved dotted line with an arrow denotes the right-lead self-energy $\Sigma_{\text{R}}$.
		(a)-(c) Diagrams for the matrix elements of electronic self-energy $\Sigma_{\text{L}\sigma,n_1n_2}(\tau_1,\tau_2)$, and the
		phononic self-energy $\overline{\Sigma}_{\text{R}\uparrow,n_1n_2}(\tau_1,\tau_2)$ and $\overline{\Sigma}_{\text{R}\downarrow,n_1n_2}(\tau_1,\tau_2)$, respectively.
		(d)-(e) Diagrams for the matrix elements of the fully interacting Green's functions $G_{\uparrow,mn}(\tau,\tau')$ and $G_{\downarrow,mn}(\tau,\tau')$, respectively.
		The filled black dots in (d) and (e) denote vertices to be integrated over.
		In the third term on the right-hand side of (d) and (e), the electron spin is flipped by phonon scattering, as indicated by the blue arrows.  }
	\label{FeynDiag}
\end{figure}

In this section, we evaluate the central-region Green's function in the presence of the self-energies of left electronic and right phononic leads, treating the electron-phonon coupling perturbatively.
Following Ref.~\cite{Funato2024}, the electron-phonon interaction between the central region and the right lead is given by
\begin{equation}\label{e-ph}
	\mathcal{H}_{e\text{-ph}} = - \sum_{m\bm{q}\lambda} \Big(
	\mathcal{J}_{\bm{q}m}   d^\dagger_{\bm{q}\lambda} 
	c^\dagger_{m\downarrow}c_{m\uparrow} + 
	\mathcal{J}_{\bm{q}m}^\ast   d_{\bm{q}\lambda} c^\dagger_{m\uparrow}c_{m\downarrow} 
	\Big) ,
\end{equation}
where $ \mathcal{J}_{\bm{q}m} $ is the electron-phonon coupling strength, and 
$d^\dagger_{\bm{q}\lambda}$ and $d_{\bm{q}\lambda}$ are defined as
\begin{equation}
	d^\dagger_{\bm{q}\lambda} =  \eta_{\bm{q}\lambda} \big(a_{\bm{q}\lambda} - a^\dagger_{-\bm{q}\bar{\lambda}} \big) ,\qquad
	d_{\bm{q}\lambda} =  \eta_{\bm{q}\lambda}^\ast \big( a^\dagger_{\bm{q}\lambda} - a_{-\bm{q}\bar{\lambda}} \big)
\end{equation}
where $\eta_{\bm{q}\lambda} = \sqrt{\hbar\omega_{\bm{q}\lambda}/(2\rho V)} \big[ (\bm{q}\times \bm{e}_{\bm{q}\lambda})_x + i (\bm{q}\times \bm{e}_{\bm{q}\lambda})_y \big]$.
The contour-ordered Green's function for a spin $\uparrow$ electron in the central region 
is defined as
\begin{equation}
	G_{\uparrow,mn}(\tau,\tau') = -\frac{i}{\hbar} \big\langle \mathcal{T}_c c_{m\uparrow}(\tau) c^\dagger_{n\uparrow}(\tau') \big\rangle 
	= -\frac{i}{\hbar} \big\langle \mathcal{T}_c \mathcal{S} c_{m\uparrow}(\tau) c^\dagger_{n\uparrow}(\tau') \big\rangle _{0} .
\end{equation}
Here $\mathcal{T}_c$ is the contour-ordering operator. The averages $\langle \cdots \rangle$ and $\langle \cdots \rangle_0$ are taken with respect to the system with and without the electron-phonon interaction, respectively. The $\mathcal{S}$-matrix is given in Eq.~(\ref{S_mat}).
Since expectation values containing an odd number of operators vanish, we retain only 
even orders in $\mathcal{J}_{\bm{q}m}$. At zeroth order
\begin{equation}
	G^{(0)}_{\uparrow,mn}(\tau,\tau') = -\frac{i}{\hbar} \big\langle \mathcal{T}_c c_{m\uparrow}(\tau) c^\dagger_{n\uparrow}(\tau') \big\rangle _0 .
\end{equation}
To second order in $\mathcal{J}_{\bm{q}m}$, we obtain
\begin{equation}
	G^{(2)}_{\uparrow,mn}(\tau,\tau') = \sum_{n_1n_2} \iint_c d\tau_1 d\tau_2 \Big[g_{\uparrow,mn_1}(\tau,\tau_1) \overline{\Sigma}^0_{\text{R}\uparrow,n_1n_2}(\tau_1,\tau_2) g_{\uparrow,n_2n}(\tau_2,\tau') \Big] ,
\end{equation}
where we have defined the following noninteracting Green’s functions
\begin{gather}
	g_{\uparrow,mn_1}(\tau,\tau_1) = -\frac{i}{\hbar} \big\langle \mathcal{T}_c c_{m\uparrow}(\tau) c^\dagger_{n_1\uparrow}(\tau_1) \big\rangle _0 , \qquad
	g_{\downarrow,n_1n_2}(\tau_1,\tau_2) = -\frac{i}{\hbar} \big\langle \mathcal{T}_c c_{n_1\downarrow}(\tau_1) c^\dagger_{n_2\downarrow}(\tau_2) \big\rangle _0 , \\
	g_{\bm{q}\lambda}(\tau_1,\tau_2) = -\frac{i}{\hbar} \big\langle \mathcal{T}_c d_{\bm{q}\lambda}(\tau_1) d^\dagger_{\bm{q}\lambda}(\tau_2) \big\rangle _0 ,\qquad
	g_{\uparrow,n_2n}(\tau_2,\tau') = -\frac{i}{\hbar} \big\langle \mathcal{T}_c c_{n_2\uparrow}(\tau_2) c^\dagger_{n\uparrow}(\tau') \big\rangle _0 ,  
\end{gather}
and  phonon self-energy
\begin{equation}\label{bornapp}
	\overline{\Sigma}^0_{\text{R}\uparrow,n_1n_2}(\tau_1,\tau_2) 
	= i\hbar \Big[  g_{\downarrow,n_1n_2}(\tau_1,\tau_2)    \sum_{\bm{q}\lambda} 
	\mathcal{J}^\ast_{\bm{q}n_1} g_{\bm{q}\lambda}(\tau_1,\tau_2) \mathcal{J}_{\bm{q}n_2} 
	\Big]
	= i\hbar \big[  g_{\downarrow,n_1n_2}(\tau_1,\tau_2) \Sigma_{\text{R},n_1n_2}(\tau_1,\tau_2)  \big] ,
\end{equation}
where the superscript 0 indicates that the phonon self-energy is evaluated using the noninteracting Green's function.
Summing all higher-order contributions and including the left-lead self-energy $\Sigma_{\text{L}\uparrow,n_1n_2}(\tau_1,\tau_2)$, we obtain the Dyson equation for $G_{\uparrow,mn}(\tau,\tau')$
\begin{equation}\label{dyson_up}
	G_{\uparrow,mn}(\tau,\tau') = g_{\uparrow,mn}(\tau,\tau') + \sum_{n_1n_2} \iint_c d\tau_1 d\tau_2 \Big[
	g_{\uparrow,mn_1}(\tau,\tau_1)   \Sigma_{\uparrow,n_1n_2}(\tau_1,\tau_2)   G_{\uparrow,n_2n}(\tau_2,\tau')\Big] ,
\end{equation}
where $\Sigma_{\uparrow,n_1n_2}(\tau_1,\tau_2) = \Sigma_{\text{L}\uparrow,n_1n_2}(\tau_1,\tau_2)  +  \overline{\Sigma}^0_{\text{R}\uparrow,n_1n_2}(\tau_1,\tau_2)$.
The left-lead electronic self-energy $\Sigma_{\text{L}\sigma}$ can be calculated numerically using an iterative algorithm~\cite{Sancho1984, Sancho1985}. 
The phonon self-energy in Eq.~(\ref{bornapp}), evaluated with the noninteracting Green's function, corresponds to the lowest-order Born approximation.
In our calculation, we first use Eq.~(\ref{bornapp}) to obtain the central-region Green’s function with Eq.~(\ref{dyson_up}), and then update the phonon self-energy by replacing the noninteracting Green’s function with the newly obtained one. This procedure is repeated until convergence is reached. This is the self-consistent Born approximation and the corresponding Feynman diagrams are illustrated in Fig.~\ref{FeynDiag}.

Using the Langreth rules \cite{Jauho2008} and Fourier transforming to energy space, the Dyson and Keldysh equations in matrix form are expressed as
\begin{gather}
	\textbf{G}^{r/a}_{\uparrow}(E)  = \textbf{g}_{\uparrow}^{r/a}(E)   +   \textbf{g}^{r/a}_{\uparrow}(E) \, \bm{\Sigma}^{r/a}_{\uparrow}(E) \, \textbf{G}^{r/a}_{\uparrow}(E)  , \\
	\textbf{G}^{>/<}_{\uparrow}(E) = \textbf{G}^{r}_{\uparrow}(E)  \,  \bm{\Sigma}^{>/<}_{\uparrow}(E) \,  \textbf{G}^{a}_{\uparrow}(E)  ,
\end{gather}	
where $\bm{\Sigma}^{>/</r/a}_{\uparrow}(E) = \bm{\Sigma}^{>/</r/a}_{\text{L}\uparrow}(E) + \overline{\bm{\Sigma}}^{>/</r/a}_{\text{R}\uparrow}(E)$.
The phonon self-energies $\overline{\bm{\Sigma}}^{>/</r/a}_{\text{R}\uparrow}(E)$ are given by
\begin{align}
	\overline{\Sigma}^{>}_{\text{R}\uparrow,ij}(E) & = i \int \frac{d\omega}{2\pi} \Big[G^{>}_{\downarrow,ij}(E-\hbar\omega)  \Sigma^{>}_{\text{R},ij}(\omega)\Big] , \\
	\overline{\Sigma}^{<}_{\text{R}\uparrow,ij}(E) & = i \int \frac{d\omega}{2\pi} \Big[G^{<}_{\downarrow,ij}(E-\hbar\omega)  \Sigma^{<}_{\text{R},ij}(\omega)\Big] , \\
	\overline{\Sigma}^{r}_{\text{R}\uparrow,ij}(E) & = i \int \frac{d\omega}{2\pi} 
	\Big[  G^{r}_{\downarrow,ij}(E-\hbar\omega)  \Sigma^{<}_{\text{R},ij}(\omega)    +    G^{>}_{\downarrow,ij}(E-\hbar\omega)  \Sigma^{r}_{\text{R},ij}(\omega)  \Big] , \\
	\overline{\Sigma}^{a}_{\text{R}\uparrow,ij}(E) & = i \int \frac{d\omega}{2\pi} 
	\Big[  G^{a}_{\downarrow,ij}(E-\hbar\omega)  \Sigma^{<}_{\text{R},ij}(\omega)    +    G^{>}_{\downarrow,ij}(E-\hbar\omega)  \Sigma^{a}_{\text{R},ij}(\omega)  \Big] .
\end{align}
Here $\textbf{G}_{\downarrow}(E-\hbar\omega)$ is the Green's function for a spin 
$\downarrow$ electron.
The subscripts $i$ and $j$ label the matrix elements.

By an analogous derivation, we obtain the Dyson and Keldysh equations for a spin 
$\downarrow$ Green's functions
\begin{gather}
	\textbf{G}^{r/a}_{\downarrow}(E)  = \textbf{g}_{\downarrow}^{r/a}(E)   +   \textbf{g}^{r/a}_{\downarrow}(E) \, \bm{\Sigma}^{r/a}_{\downarrow}(E) \, \textbf{G}^{r/a}_{\downarrow}(E)  , \\
	\textbf{G}^{>/<}_{\downarrow}(E) = \textbf{G}^{r}_{\downarrow}(E)  \,  \bm{\Sigma}^{>/<}_{\downarrow}(E) \,  \textbf{G}^{a}_{\downarrow}(E)  ,
\end{gather}	
where $\bm{\Sigma}^{>/</r/a}_{\downarrow}(E) = \bm{\Sigma}^{>/</r/a}_{\text{L}\downarrow}(E) + \overline{\bm{\Sigma}}^{>/</r/a}_{\text{R}\downarrow}(E)$.
The phonon self-energies $\overline{\bm{\Sigma}}^{>/</r/a}_{\text{R}\downarrow}(E)$ are given by
\begin{align}
	\overline{\Sigma}^{>}_{\text{R}\downarrow,ij}(E) & = i \int \frac{d\omega}{2\pi} \Big[G^{>}_{\uparrow,ij}(E+\hbar\omega)  \Sigma^{<}_{\text{R},ji}(\omega)\Big] ,\\
	\overline{\Sigma}^{<}_{\text{R}\downarrow,ij}(E) & = i \int \frac{d\omega}{2\pi} \Big[G^{<}_{\uparrow,ij}(E+\hbar\omega)  \Sigma^{>}_{\text{R},ji}(\omega)\Big] ,\\
	\overline{\Sigma}^{r}_{\text{R}\downarrow,ij}(E) & = i \int \frac{d\omega}{2\pi} \Big[G^{r}_{\uparrow,ij}(E+\hbar\omega)  \Sigma^{<}_{\text{R},ji}(\omega) + G^{<}_{\uparrow,ij}(E+\hbar\omega)  \Sigma^{a}_{\text{R},ji}(\omega)  \Big] ,\\
	\overline{\Sigma}^{a}_{\text{R}\downarrow,ij}(E) & = i \int \frac{d\omega}{2\pi} \Big[G^{a}_{\uparrow,ij}(E+\hbar\omega)  \Sigma^{<}_{\text{R},ji}(\omega) + G^{<}_{\uparrow,ij}(E+\hbar\omega)  \Sigma^{r}_{\text{R},ji}(\omega)  \Big] .
\end{align}
The relations between the left-lead self-energy $\bm{\Sigma}_{\text{L}\sigma}(E)$ and the 
corresponding linewidth function $\bm{\Gamma}_{\text{L}\sigma}(E)$ are \cite{Jauho2008}
\begin{equation}
	\bm{\Sigma}^{>}_{\text{L}\sigma}(E) = i \big[ f_{\text{L}}(E) - 1 \big] \bm{\Gamma}_{\text{L}\sigma}(E) ,\qquad
	\bm{\Sigma}^{<}_{\text{L}\sigma}(E) = i f_{\text{L}}(E) \bm{\Gamma}_{\text{L}\sigma}(E) ,\qquad
	\bm{\Gamma}_{\text{L}\sigma}(E) = i \big[ \bm{\Sigma}^{r}_{\text{L}\sigma}(E) - \bm{\Sigma}^{a}_{\text{L}\sigma}(E) \big],
\end{equation}
where $f_{\text{L}}(E) = \big[e^{(E-\mu_\text{L})/(k_{\text{B}}T_\text{L})} + 1\big]^{-1}$ is the Fermi–Dirac distribution.
We model the right phonon reservoir with a super-Ohmic spectral density~\cite{Weissbook, Thorwart2005, TQS2022}
\begin{equation}\label{superOm}
	J_{\text{R}}(\omega) = \alpha \hbar\frac{\omega^3}{\omega_0^{2}} e^{-\omega/\omega_0} ,
\end{equation}
where the dimensionless parameter $\alpha$ describes the dissipation into the reservoir and $\hbar\omega_0$ is the cutoff energy.
The right-lead linewidth function is taken to be $\bm{\Gamma}_{\text{R}}(\omega) = 
J_{\text{R}}(\omega)\textbf{I}$.
The relations between the right-lead self-energy $\bm{\Sigma}_{\text{R}}(\omega)$ and $\bm{\Gamma}_{\text{R}}(\omega)$ are~\cite{Jauho2008}
\begin{equation}
	\bm{\Sigma}_\text{R}^>(\omega) = -i \big[ n_\text{R}(\omega) + 1 \big] \bm{\Gamma}_\text{R}(\omega)  , \qquad
	\bm{\Sigma}_\text{R}^<(\omega) = -i n_\text{R}(\omega) \bm{\Gamma}_\text{R}(\omega)  , \qquad
	\bm{\Sigma}_\text{R}^r(\omega) = \big[ \bm{\Sigma}_\text{R}^a(\omega) \big]^\dagger = -i\bm{\Gamma}_\text{R}(\omega) /2 ,
\end{equation}
where $n_\text{R}(\omega) = \big[e^{(\hbar\omega)/(k_{\text{B}}T_\text{R})} - 1\big]^{-1}$ is the Bose–Einstein distribution.
Using Eq.~(\ref{Is1}) and Eq.~(\ref{I_e}), the spin current in the left lead is expressed as
\begin{align}\label{Is2}
	I_{\text{s}} & = \frac{\hbar}{2e} \left( I^e_{\uparrow} - I^e_{\downarrow} \right)   
	=   \frac{\hbar}{2e}  \left( I^e_{\uparrow} + I^e_{\uparrow} \right)  
	=  \frac{\hbar}{e}  I^e_{\uparrow}  \nonumber\\
	& = \int \frac{dE}{2\pi} \text{Tr} \left\{ 
	\textbf{G}^r_{\uparrow}(E) \left[if_{\text{L}}(E) \overline{\bm{\Sigma}}^{>}_{\text{R}\uparrow}(E) + i\big(1-f_\text{L}(E)\big) \overline{\bm{\Sigma}}^{<}_{\text{R}\uparrow}(E) \right] \textbf{G}^a_{\uparrow}(E) \bm{\Gamma}_{\text{L}\uparrow}(E) \right\} .
\end{align}
For an insulating electronic system, $I^e_{\uparrow} = - I^e_{\downarrow}$ and therefore, for simplicity, we consider only the spin-up contribution here.

\section{S3. The expression of the phonon angular momentum current in the right lead}\label{Is_R}
In what follows, we derive the expression of the phAM current in the right lead. The phonon number operator of the right lead is expressed as $N_{\text{R}} = \sum_{\bm{q}\lambda} a^\dagger_{\bm{q}\lambda} a_{\bm{q}\lambda}$,
where $\lambda=\pm$ labels the phonon chirality. We define $\mathcal{N}_{\text{R}}$ as the difference between the phonon numbers of opposite chiralities as
\begin{equation}
	\mathcal{N}_{\text{R}} = \sum_{\bm{q}} \big( a^\dagger_{\bm{q}+} a_{\bm{q}+} - a^\dagger_{\bm{q}-} a_{\bm{q}-} \big) .
\end{equation}
The phAM current is
\begin{equation}\label{I_ph}
	I_{\text{ph}} = \hbar \Big\langle \frac{d}{dt} \mathcal{N}_{\text{R}} \Big\rangle
	= \hbar \sum_{m} \big\{ \big[ G^{<}_{m,+}(t,t) - G^{<}_{m,-}(t,t) \big]  +  \text{h.c.}  \big\} ,
\end{equation}
where the lesser Green's functions are defined as
\begin{equation}
	G^<_{m,+}(t,t') = -\frac{i}{\hbar}  \sum_{\bm{q}} \mathcal{J}_{\bm{q}m}     
	\big\langle  d^\dagger_{\bm{q}+}(t')    c^\dagger_{m\downarrow}(t) c_{m\uparrow}(t)  
	\big\rangle , \qquad
	G^<_{m,-}(t,t') = -\frac{i}{\hbar}  \sum_{\bm{q}} \mathcal{J}_{\bm{q}m}     
	\big\langle  d^\dagger_{\bm{q}-}(t')    c^\dagger_{m\downarrow}(t) c_{m\uparrow}(t)  
	\big\rangle .
\end{equation}
We then introduce the corresponding contour-ordered Green’s functions
\begin{align}
	G_{m, +}(\tau, \tau') &	= -\frac{i}{\hbar} \sum_{\bm{q}} \mathcal{J}_{\bm{q}m}  
	\big\langle \mathcal{T}_c \mathcal{S} c^\dagger_{m\downarrow}(\tau) 
	c_{m\uparrow}(\tau) d^\dagger_{\bm{q}+}(\tau') \big\rangle_0 , \\
	G_{m, -}(\tau, \tau') & = -\frac{i}{\hbar} \sum_{\bm{q}} \mathcal{J}_{\bm{q}m}  
	\big\langle \mathcal{T}_c \mathcal{S} c^\dagger_{m\downarrow}(\tau) 
	c_{m\uparrow}(\tau) d^\dagger_{\bm{q}-}(\tau') \big\rangle_0 .
\end{align}
Using a similar procedure for the perturbative expansion as that used in the previous section, we obtain
\begin{align}
	G^{<}_{m,\pm}(t,t) & = G_{m, +}(t,t) - G_{m, -}(t,t)  \nonumber\\
	& = \frac{1}{\hbar} \sum_{m_1} \int \frac{dE}{2\pi} 
	\big[ g^{r}_{\uparrow,mm_1}(E) \overline{\Sigma}^{<}_{\text{R},m_1m}(E)  +  g^{<}_{\uparrow,mm_1}(E) \overline{\Sigma}^{a}_{\text{R},m_1m}(E)  \big] 
\end{align}
where
\begin{align}
	\overline{\Sigma}^{<}_{\text{R},m_1m}(E) & = i \int \frac{d\omega}{2\pi} \big[ g^<_{\downarrow,m_1m}(E-\hbar\omega) \Sigma^<_{\text{R},m_1m}(\omega) \big] , \\
	\overline{\Sigma}^{a}_{\text{R},m_1m}(E) & = i \int \frac{d\omega}{2\pi} \big[ g^a_{\downarrow,m_1m}(E-\hbar\omega) \Sigma^<_{\text{R},m_1m}(\omega)  +  g^>_{\downarrow,m_1m}(E-\hbar\omega) \Sigma^a_{\text{R},m_1m}(\omega) \big] .
\end{align}
Likewise, we model the right phonon reservoir with a super-Ohmic spectral density given in Eq.~(\ref{superOm}).
In the end, the phAM current is
\begin{equation}
	I_{\text{ph}}  = \int \frac{dE}{2\pi} \text{Tr} \big\{ \textbf{G}^{r}_{\uparrow}(E)  \bm{\Gamma}_{\text{L}\uparrow}(E)  \textbf{G}^{a}_{\uparrow}(E)  
	\big[2f_{\text{L}}(E) \text{Im}\overline{\bm{\Sigma}}^{r}_{\text{R}}(E) - i \overline{\bm{\Sigma}}^{<}_{\text{R}}(E) \big] \big\} .
\end{equation}

\section{S4. Verification of current conservation}\label{cc}
\begin{figure}[h]
	\centering
	\includegraphics[scale=0.32]{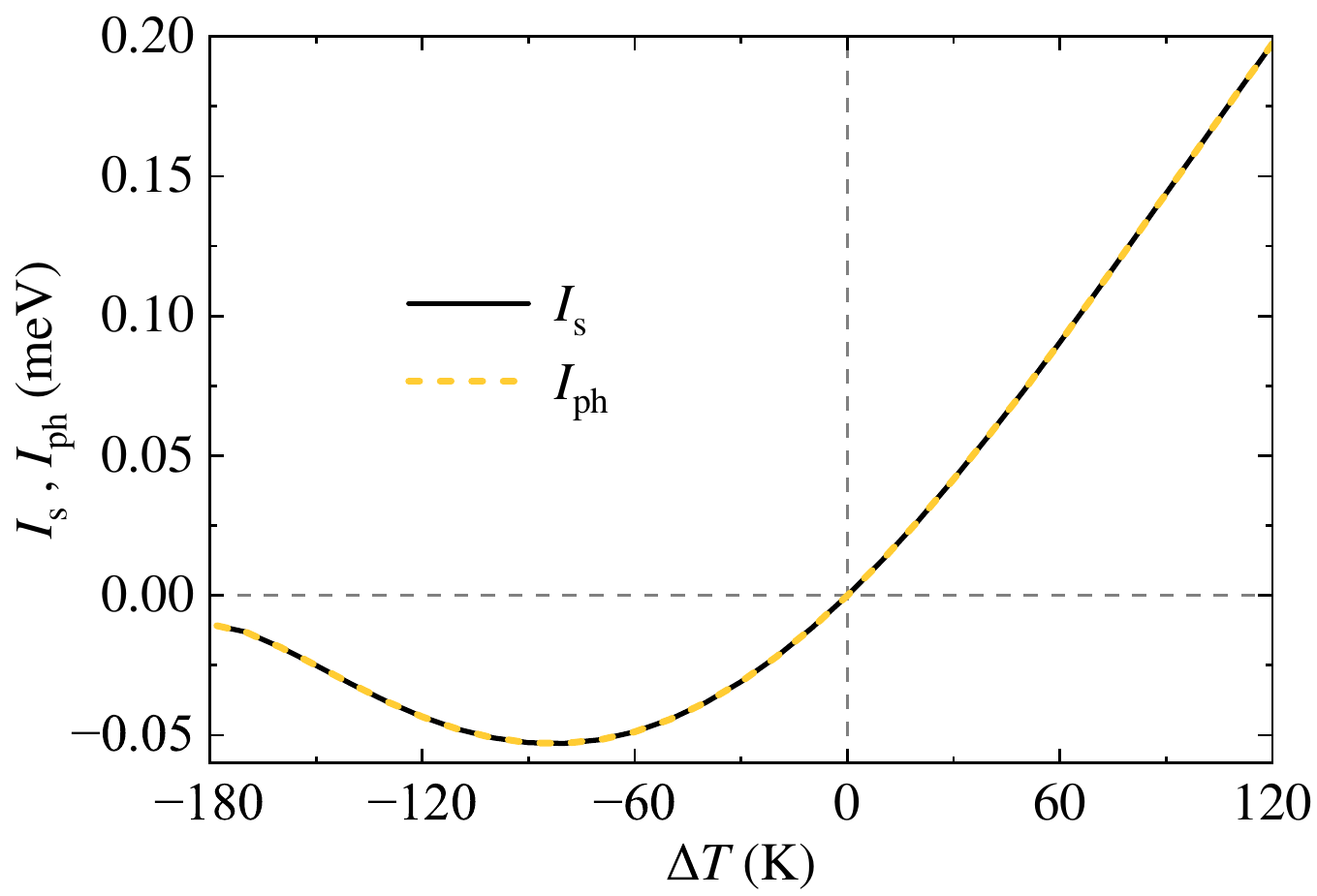}
	\caption{
		Spin current $I_{\text{s}}$ (blue solid line) and phAM current $I_{\text{ph}}$ (orange dotted line) plotted as functions of the temperature difference $\Delta T$.
		Other parameters: $T_0=90\,\rm K$, $\mu_{\text{L}}=1\,\rm meV$, $\alpha=5$ and $\hbar\omega_0=7\,\rm meV$.}
	\label{conservation}
\end{figure}

The spin current is usually not conserved in the presence of the spin-orbit interaction 
(SOI) ~\cite{WJ06, SQF05, SJR06}.
For simplicity, we neglect the SOI-induced spin nonconservation in the NM,
since our primary focus is the generation of spin current via spin-phonon conversion.
Accordingly, the spin current $I_{\text{s}}$ in the left lead should be equal in magnitude to the phonon angular momentum current $I_{\text{ph}}$ in the right lead, i.e., $I_{\text{s}} = I_{\text{ph}}$.
The current conservation is verified numerically in Fig.~\ref{conservation}.

\section{S5. The lattice Hamiltonian of the normal metal}\label{NM}
In this section, we present the tight-binding Hamiltonian of the normal metal.
We focus on the 3D case; the 1D and 2D cases can be obtained by a similar procedure.
We begin with the Hamiltonian with only the kinetic energy of $N_e$ electrons
\begin{equation}
	H_{3\text{D}} = \sum_{i=1}^{N_e} \frac{\hat{\bm{p}}_i^2}{2m} = -\frac{\hbar^2}{2m} \sum_{i=1}^{N_e} \nabla^2_i .
\end{equation}
The single-electron wave function is $\phi_{\bm{k}\sigma}(\bm{r}) = (1/\sqrt{V}) \, e^{i\bm{k}\cdot\bm{r}} |\sigma \rangle$,
where $\bm{k}$ is the wave vector and $\sigma$ denotes the spin. The Hamiltonian in second-quantized form is
\begin{equation}
	H_{3\text{D}} = \sum_{\bm{k}\sigma} \sum_{\bm{k}'\sigma'} \langle \bm{k}'\sigma'| \frac{\hat{\bm{p}}^2}{2m} | \bm{k}\sigma \rangle 
	c^\dagger_{\bm{k}'\sigma'}c_{\bm{k}\sigma}
	= \sum_{\bm{k}\sigma} \frac{\hbar^2k^2}{2m} c^\dagger_{\bm{k}\sigma}c_{\bm{k}\sigma} ,
\end{equation}
Considering the replacement $k_i^2 \rightarrow (2/a^2) \big[1-\cos(k_ia)\big] $ and performing the Fourier transform of the creation and annihilation operators with
\begin{equation}
	c_{\bm{k}\sigma} = \frac{1}{\sqrt{V}} \sum_i e^{i\bm{k}\cdot\bm{r}_i} c_{i\sigma}, \qquad \qquad
	c^\dagger_{\bm{k}\sigma} = \frac{1}{\sqrt{V}} \sum_i e^{-i\bm{k}\cdot\bm{r}_i} c^\dagger_{i\sigma},
\end{equation}
we obtain
\begin{equation}\label{H_3D}
	H_{3\text{D}} = 6t\sum_{i\sigma} c^\dagger_{i\sigma}c_{i\sigma} - t\sum_{\langle ij \rangle\sigma} c^\dagger_{i\sigma}c_{j\sigma} ,
\end{equation}
where we have defined $t = \hbar^2 / (2ma^2)$.
Here $i$ labels lattice sites and ${\langle ij \rangle}$ denotes nearest-neighbor hopping.
For 1D and 2D cases, we have
\begin{align}
	H_{1\text{D}} & = 2t\sum_{i\sigma} c^\dagger_{i\sigma}c_{i\sigma} - t\sum_{\langle ij \rangle \sigma} c^\dagger_{i\sigma}c_{j\sigma} , \label{H_1D} \\
	H_{2\text{D}} & = 4t\sum_{i\sigma} c^\dagger_{i\sigma}c_{i\sigma} - t\sum_{\langle ij \rangle \sigma} c^\dagger_{i\sigma}c_{j\sigma} . \label{H_2D}
\end{align}
In this work, both the left lead and the central region are modeled as a normal metal 
described by the tight-binding Hamiltonian above with the hopping parameter $t = 21.768 
\, \text{meV}$.

\section{S6. The spin current for different central-region sizes and 
dimensionalities}\label{dimen}
\begin{figure}[h]
	\centering
	\includegraphics[scale=0.53]{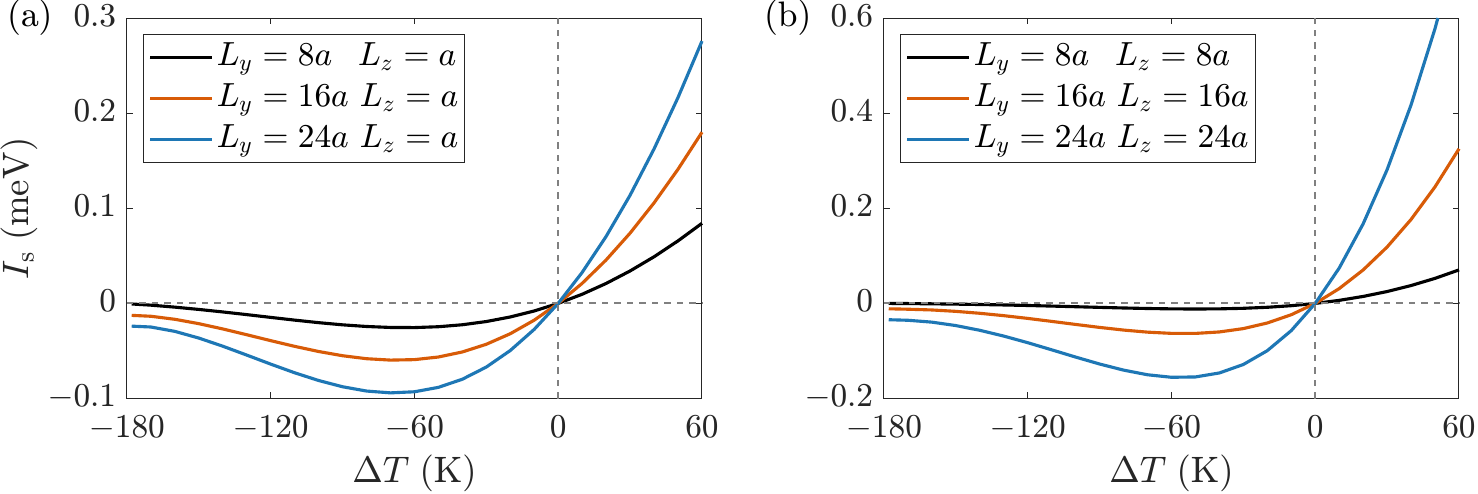}
	\caption{Spin current $I_\text{s}$ as a function of the temperature difference 
	$\Delta T$ for different central-region sizes. The width of the central region along 
	the transport direction is fixed at $L_x=a$, where $a=5\,\text{nm}$ is the lattice 
	constant. Other parameters are $T_0=90\,\rm K$, $\alpha=5$, and $\hbar\omega_0=7\,\rm 
	meV$. The chemical potential of the left lead is set to (a) 
	$\mu_{\text{L}}=3\,\text{meV}$ for the 2D model and (b) 
	$\mu_{\text{L}}=6\,\text{meV}$ for the 3D model.}
	\label{Is_size}
\end{figure}

In the main text, we model the central region as a quantum dot for simplicity.
Here we present the calculated spin current for different central-region sizes using the 2D and 3D lattice Hamiltonians given in Eqs.\,\eqref{H_2D} and \eqref{H_3D}, respectively.
The results, shown in Figs.\,\ref{Is_size}(a)-\ref{Is_size}(b), demonstrate that the negative differential SSE also persists in 2D and 3D systems, although the magnitude of the spin current depends on the dimensionality and the central-region size.

\bibliography{reference_sm}